\documentclass[aps,prl,reprint,groupedaddress,showpacs,showkeys,nofootinbib,amsmath,
amssymb,amsfonts]{revtex4-1}
\usepackage{color}
\usepackage{graphicx}
\usepackage{dcolumn}
\usepackage{bm}
\usepackage[bookmarks=true,colorlinks,
            citecolor=blue,linkcolor=blue,urlcolor=blue,
            breaklinks=true]{hyperref}

\begin{document}
\title{All Order Linearized Hydrodynamics from Fluid/Gravity Correspondence}
\date{\today}
\author{Yanyan Bu}
\email{yybu@post.bgu.ac.il}
\author{Michael Lublinsky}
\email{lublinm@bgu.ac.il}
\affiliation{Department of Physics, Ben-Gurion University of the Negev, Beer-Sheva 84105, Israel}

\begin{abstract}
Using fluid/gravity correspondence, we determine the (linearized) stress energy tensor of $\mathcal{N}=4$ super-Yang-Mills theory at strong coupling with all orders in derivatives of fluid velocity  included. We find that the dissipative effects are fully encoded in the shear term and a new one, which emerges  starting from the third order. We derive, for the first time, closed linear holographic RG flow-type equations for (generalized) momenta-dependent viscosity functions. In the hydrodynamic regime, we obtain the stress tensor up to third order in derivative expansion analytically. We then numerically determine the viscosity functions up to large momenta. As a check of our results, we also derive the generalized Navier-Stokes equations from the Einstein equations in the dual gravity.
\end{abstract}
\maketitle
\section{Introduction}
The quark-gluon plasma produced in heavy ion collisions behaves like a nearly perfect fluid reflecting strongly coupled regime of QCD \cite{nucl-th/0405013,hep-ph/0405066}. Relativistic hydrodynamics is found to describe QCD plasma expansion near thermal equilibrium. Meanwhile, various microscopic models are indispensable in order to understand the transport properties of this fascinating QCD matter.

An important tool to address the strongly coupled dynamics is the AdS/CFT correspondence \cite{hep-th/9711200}, which reformulates the large $N$ strongly interacting quantum field theory in terms of classical gravity in  (asymptotically) AdS spacetime. One celebrated prediction of the AdS/CFT correspondence is the universal value of shear viscosity over entropy density ratio \cite{hep-th/0104066,hep-th/0405231}, valid for a large class of large $N$ strongly coupled gauge theory plasmas which have Einstein gravity duals. Subsequent exploration of fluid dynamics from  gravity in  AdS black hole geometries has become a major research topic, see Ref.~\cite{0704.0240} for a review and references therein.

The authors of Refs.~\cite{0704.1647,0905.4069} (see also Refs.~\cite{1103.3452,1203.0755,1302.0697}) proposed a new generalized relativistic hydrodynamics with all orders in derivatives of fluid velocity resummed in the stress tensor. Higher order derivatives can be classified into non-linear (like $(\nabla u)^2$) and linear terms (like $\nabla \nabla u$) with respect to the local fluid velocity. The non-linearities are important when the velocity field amplitude is large. However, even for small amplitude waves, one can get large contributions from the linear terms when the momenta associated with the wave are large. In~\cite{0905.4069}, only linear terms were kept in the stress tensor. Viscosity and other constant value transport coefficients  were generalized into  momenta dependent functions that collected all higher order terms in a self-consistent manner. This viscosity function is expressed in momentum space which follows from the replacement $\partial_{\mu}\to \left(-i\omega,i\vec{q}\right)$ in the linear gradient expansion of stress tensor $T_{\mu\nu}$. With constitutive relations at hand, these transport coefficient functions were supposed to be deduced  from retarded correlators \cite{hep-th/0205051,hep-th/0602059} computed in linearized bulk gravity. However, a generic problem prevented achieving this goal: knowledge of the retarded correlators happened to be insufficient to determine all the transport coefficient functions. This paper reports on a major progress in generalizing relativistic hydrodynamics to all orders, as envisioned in~\cite{0704.1647,0905.4069}:
We consistently derive the transport coefficient functions by extending~\cite{0712.2456,0806.0006,0809.4272} to linearized fluid/gravity correspondence.

The fluid/gravity correspondence maps construction of fluid stress tensor and its conservation law (Navier-Stokes equations) into the problem of solving Einstein equations in asymptotically AdS spacetime. In particular, it provides  a systematic framework to study non-linear fluid dynamics, order by order in the boundary derivative expansion. In principle, the perturbative calculations of Ref.~\cite{0712.2456} can be extended to arbitrary order in terms of derivative expansion. Our procedure is, however, somewhat different from that of~\cite{0712.2456}. We will collect the dissipative contributions\footnote{While some of the high order derivative terms might not necessarily lead to actual dissipation in the system, we collectively refer to all of them as dissipative contributions.} to the stress tensor in a unified way, rather than appealing to an order by order derivative expansion.

Our major new result is closed holographic RG flow-type equations derived for the viscosity functions. These are linear equations, which we first study analytically using  perturbative expansion and then exactly numerically, leading to new understanding of dissipative structures of strongly coupled plasmas.
\section{Linearized fluid/gravity correspondence}
We consider the universal sector of the AdS/CFT correspondence: the dynamics of Einstein gravity with a negative cosmological constant in five dimensional spacetime,
\begin{equation}\label{action}
S=\frac{1}{16\pi G_{N}}\int d^5x\sqrt{-g}(R+12).
\end{equation}
The 4-parameter family of solutions to action~(\ref{action}) is
\begin{equation}\label{boosted bh}
ds^2=-2u_{\mu}dx^{\mu}dr+r^2\left(\mathcal{P}_{\mu\nu}-f\left({\bf b} r\right) u_{\mu}u_{\nu} \right)dx^{\mu}dx^{\nu},
\end{equation}
with
\begin{equation}\label{velocity}
u_v=-\frac{1}{\sqrt{1-\beta^2}},~~~u_i=\frac{\beta_{i}}{\sqrt{1-\beta^2}},~~~ \beta^2=\sum_{i=1}^{3}\beta_i\beta_i,
\end{equation}
and the function $f(r)=1-1/r^4$. We use notation $x^{\mu}=(v,x^i)$ with $v$ to denote time direction in the Eddington-Finkelstein coordinate. Indeed, $v$ is equivalent to time in Poincare patch as $r\to\infty$ and therefore is identified as time of the boundary field theory. The Hawking temperature of the above black hole is
\begin{equation}
T=\frac{1}{\pi {\bf b}},
\end{equation}
which will be identified as the temperature of the dual fluid defined in the boundary. The projection operator $\mathcal{P}_{\mu\nu}=\eta_{\mu\nu}+u_{\mu}u_{\nu}$. Note that the parameters $\beta_i$ and ${\bf b}$ are all constants so that the line element~(\ref{boosted bh}) does form a class of solutions to the bulk Einstein equation
\begin{equation}\label{einstein eq}
E_{MN}=R_{MN}-\frac{1}{2}g_{MN}R-6g_{MN}=0.
\end{equation}

To discuss hydrodynamics, we follow Ref.~\cite{0712.2456} and promote the constant parameters $\beta_i$ and ${\bf b}$ to slowly but otherwise arbitrarily varying functions of the boundary coordinates,
\begin{equation}\label{metric1}
\begin{split}
ds^2=&-2u_{\mu}(x^{\alpha})dx^{\mu}dr +r^2 \mathcal{P}_{\mu\nu}(x^{\alpha}) dx^{\mu}dx^{\nu}\\
&-r^2f\left({\bf b}(x^{\alpha}) r\right) u_{\mu}(x^{\alpha}) u_{\nu}(x^{\alpha})dx^{\mu}dx^{\nu}.
\end{split}
\end{equation}
The metric~(\ref{metric1}) no longer solves the bulk equation~(\ref{einstein eq}). The idea of Ref.~\cite{0712.2456} is to add suitable corrections in~(\ref{metric1}) so that the bulk  equation~(\ref{einstein eq}) is satisfied by the new metric. Ref.~\cite{0712.2456} introduced a systematic way to construct the corrected metric:
the method is to first perform a boundary derivative expansion for $\beta_i(x^{\alpha})$ and ${\bf b}(x^{\alpha})$ around a chosen point, such as the origin $x^{\alpha}=0$, and then solve linearized Einstein equations in the bulk order by order in the derivative expansion.

Our goal is to sum all higher order terms in the fluid stress tensor. In contrast to the boundary derivative expansion of Ref.~\cite{0712.2456}, we  linearize the fluid fields $u_{\mu}(x^{\alpha})$ and ${\bf b}(x^{\alpha})$. We then determine the corrected metric by solving the bulk equations. Our corrected metric accounts for all order dissipative contributions to the fluid stress tensor.

The fluid velocity and temperature  are expanded as
\begin{equation}\label{linearization}
\begin{split}
u_{\mu}(x^{\alpha})&=\left(-1,\epsilon\beta_i(x^{\alpha})\right)+\mathcal{O}(\epsilon^2),\\
{\bf b}(x^{\alpha})&={\bf b}_0+\epsilon {\bf b}_1(x^{\alpha})+\mathcal{O}(\epsilon^2),
\end{split}
\end{equation}
where, as in Ref.~\cite{0712.2456}, we multiply $\beta_i$ and ${\bf b}_1$ by a small  $\epsilon$, which will be set to one in the final expression of the fluid stress tensor. ${\bf b}_0$ denotes temperature of the fluid in equilibrium while the linear term ${\bf b}_1(x^{\alpha})$ accounts for dissipative corrections. Below, ${\bf b}_0$ is set to one by conformal invariance.

In accordance with~(\ref{linearization}), the seed metric, i.e., the linearized version of~(\ref{metric1}) is,
\begin{equation}\label{line element}
\begin{split}
ds^2=&2drdv-r^2f(r)dv^2+r^2\delta_{ij}dx^idx^j\\
&-\epsilon\left[2\beta_i(x^{\alpha}) drdx^i+2r^{-2}\beta_i(x^{\alpha}) dvdx^i\right.\\
&+\left.4r^{-2} {\bf b}_1(x^{\alpha})dv^2\right]+\mathcal{O}(\epsilon^2),
\end{split}
\end{equation}
where the first line is exactly the line element of the Schwarzschild-$AdS_5$ black brane written in the ingoing Eddington-Finkelstein coordinate. The terms linear in $\epsilon$ are parts of the metric corrections we are after. We are to introduce metric corrections up to $\mathcal{O}(\epsilon)$. The full metric is
\begin{equation}
g=g^{(0)}+\epsilon g^{(1)} \left[\beta_i,{\bf b}_1\right]+ \mathcal{O}(\epsilon^2),
\end{equation}
where $g^{(0)}$ is the first line of~(\ref{line element}). The first order correction $g^{(1)}$ has two sources: the first one is already known, corresponding to linear terms in $\epsilon$ in~(\ref{line element}); while the second contribution will be determined from the bulk dynamics and summarized in the line element~(\ref{line element correction}).

Diffeomorphism invariance allows us to choose a gauge. Following ~\cite{0712.2456}, we  work in the ``background field'' gauge,
\begin{equation}\label{bk gauge}
g_{rr}=0,~~~g_{r\mu}\propto u_{\mu},~~~\textrm{Tr}\left[(g^{(0)})^{-1}g^{(1)}\right]=0.
\end{equation}
The line element for the undetermined metric is~\cite{0712.2456},
\begin{equation}\label{line element correction}
\begin{split}
ds^2_{(1)}=\epsilon\left[-3hdrdv+r^{-2}kdv^2 +r^2 h\delta_{ij}dx^idx^j\right.\\
\left.+2r^2(1-f(r))j_{i}dvdx^i+r^2\alpha_{ij}dx^idx^j\right],
\end{split}
\end{equation}
where $\alpha_{ij}$ is a symmetric traceless tensor of rank two. All the components $\{h,~k,~j_{i},~\alpha_{ij}\}$ are explicit functions of the bulk coordinates $\{x^{\alpha},r\}$. Their precise forms have to be determined from  equation~(\ref{einstein eq}), supplemented with proper boundary conditions to be discussed next.

The first boundary condition is regularity requirement for all the components over the whole range of $r$, in particular at the unperturbed horizon $r=1$. This is a natural choice since the ingoing Eddington-Finkelstein coordinate we are working in is free of coordinate singularity. The second boundary condition comes from the asymptotic consideration at $r\to \infty$. Since the dual fluid is in Minkowski space with metric $\eta_{\mu\nu}$, we will require that the metric corrections should not change the asymptotic behavior  of the metric~(\ref{metric1}). The latter condition strongly constrains the large $r$ behavior of different components $\{h,~k,~j_{i},~\alpha_{ij}\}$: as $r\to \infty$, their falling-off behaviors should be restricted as
\begin{equation}\label{AdS constraint}
h< \mathcal{O}(1),~~~k<\mathcal{O}(r^4),~~~j_i<\mathcal{O}(r^4),~~~ \alpha_{ij}<\mathcal{O}(1).
\end{equation}
Some of the integration constants  remain unfixed by the above considerations. This is due to the freedom of defining  fluid velocity. This ambiguity will be removed by appropriately choosing a frame for the dual fluid. To be specific, we will work in the ``Landau frame''
\begin{equation}\label{frame convention}
u^{\mu}T_{\mu\nu}^{\text{Diss}}=0,
\end{equation}
where $T_{\mu\nu}^{\text{Diss}}$ is the dissipative part of $T_{\mu\nu}$.

Once the dual metric is worked out, the stress tensor of the dual fluid is  calculated from the formula~\cite{hep-th/9902121,hep-th/9806087}
\begin{equation}\label{stress tensor1}
T_{\nu}^{\mu}=-\lim_{r\rightarrow \infty}r^4\left(2\mathcal{K}_{\nu}^{\mu}-
2\mathcal{K}\gamma_{\nu}^{\mu}+ 6\gamma_{\nu}^{\mu} -G^{\mu}_{\nu}\right),
\end{equation}
where $\gamma_{\mu\nu}$ and $\mathcal{K}_{\mu\nu}$ are the induced metric and extrinsic curvature on the hypersurface with fixed $r$, respectively. The Einstein tensor $G_{\mu\nu}$ is compatible with $\gamma_{\mu\nu}$.

\section{Fluid dynamics dual to bulk gravity}
We are now to study the bulk dynamics. The Einstein equations are divided into dynamical and constraints. Our strategy is to first solve the former. This will lead to a construction of an ``off-shell" boundary stress tensor. The constraints will be later shown to be equivalent to the stress tensor conservation. We start from $E_{rr}=0$,
\begin{equation}\label{h eq}
5\partial_rh+r\partial_r^2h=0,
\end{equation}
which is  Eq.~(4.7) of Ref.~\cite{0712.2456}. 
Generic solution  is
\begin{equation}
h(x^{\alpha},r)=s_0(x^{\alpha})+s_1(x^{\alpha})~r^{-4},
\end{equation}
where $s_0$ and $s_{1}$ are arbitrary functions of boundary coordinates $x^{\alpha}$. A nonzero function $s_0$ violates the asymptotic requirement for $h$ as specified in Eq.~(\ref{AdS constraint}). In addition, $s_1\neq 0$ is equivalent to $T^{\text{Diss}}_{00}\neq 0$. Therefore, the constraint from asymptotic infinity and ``Landau frame'' convention lead to $h=0$.

The function $k$ will be found from $E_{rv}=0$,
\begin{equation}\label{k eq}
3r^2\partial_r k=6r^4\partial\beta+r^3\partial_v\partial\beta-2\partial j- r\partial_r\partial j-r^3\partial_i\partial_j\alpha_{ij},
\end{equation}
where $\partial j\equiv\partial^i j_i$. The scalar function $k$ cannot be determined until $j_i$ and $\alpha_{ij}$ are found. Fortunately, the  dynamical equations for $j_i$ and $\alpha_{ij}$ are not entangled with $k$, so we   integrate  Eq.~(\ref{k eq}) after solving for
$j_i$ and $\alpha_{ij}$.

In order to determine $j_i$, we consider $E_{ri}=0$,
\begin{equation}\label{ji eq}
\partial_r^2 j_i=\left(\partial_i\partial\beta-\partial^2\beta_i\right)-3r \partial_v \beta_i+\frac{3}{r}\partial_r j_i -r^2\partial_r\partial_j\alpha_{ij},
\end{equation}
where $\partial^2\equiv\partial_i\partial^i$. To find $\alpha_{ij}$ is more involved as its diagonal and non-diagonal components have to be treated separately. Here we   report the final result,
\begin{equation}\label{alphaij eq}
\begin{split}
0=&(r^7-r^3)\partial_r^2 \alpha_{ij}+(5r^6-r^2)\partial_r\alpha_{ij}\\
&+2r^5\partial_v\partial_r\alpha_{ij}+ 3r^4\partial_v\alpha_{ij}+ r^3[[\alpha_{ij}]]\\
&+\left(1-r\partial_r\right)[[j]] + 2\left(3r^4+r^3\partial_v\right)\sigma_{ij},
\end{split}
\end{equation}
with $\sigma_{ij}=\frac{1}{2}\left(\partial_i\beta_j+\partial_j\beta_i-\frac{2}{3} \delta_{ij} \partial\beta\right)$. Two functionals were introduced in Eq.~(\ref{alphaij eq}),
\begin{equation}\nonumber
\begin{split}
[[\alpha_{ij}]]&\equiv\partial^2\alpha_{ij}-\left(\partial_i\partial_k\alpha_{jk} +
\partial_j\partial_k\alpha_{ik}-\frac{2}{3}\delta_{ij} \partial_k\partial_l\alpha_{kl}\right),\\
[[j]]&\equiv\partial_i j_j+\partial_j j_i-\frac{2}{3}\delta_{ij} \partial j.
\end{split}
\end{equation}

$j_i$ and $\alpha_{ij}$ are uniquely decomposed as
\begin{equation}\label{decomposition}
\left\{
\begin{aligned}
j_i=&a(\omega,q,r)\beta_i+b(\omega,q,r)\partial_i\partial\beta,\\
\alpha_{ij}=&2c(\omega,q,r)\sigma_{ij}+ d(\omega,q,r)\pi_{ij},
\end{aligned}
\right.
\end{equation}
where $\pi_{ij}=\partial_i\partial_j\partial\beta-\frac{1}{3}\delta_{ij} \partial^2\partial\beta$. The decomposition~(\ref{decomposition}) is inspired by the source terms in~(\ref{ji eq}) and~(\ref{alphaij eq}). On the one hand, the homogeneous part of the solutions for~(\ref{ji eq}) and~(\ref{alphaij eq}) did not appear in the above decomposition due to the ``Landau frame'' convention~(\ref{frame convention}) and the large $r$ requirement~(\ref{AdS constraint}). On the other hand, since ${\bf b}_1$ does not contribute in the source terms of~(\ref{ji eq}) and~(\ref{alphaij eq}), we do not consider derivatives of ${\bf b}_1$ as basis vector/tensor in~(\ref{decomposition}). Explicit calculations can be done to show that adding derivatives of ${\bf b}_1$ in the above decomposition will result in similar equations as~(\ref{abcd eqs}) but without source terms. Then, the boundary conditions force these added modes to vanish.
For convenience, we prefer to express the coefficient functions in momentum space but with tensors $\sigma_{ij}$ and $\pi_{ij}$ formulated as explicit derivatives of the fluid velocity. The momentum variables are in one to one correspondence with derivative operators in accord with the replacement rule $\partial_{\mu}\rightarrow \left(-i\omega,~i\vec{q}\right)$. We are led to a system of ordinary differential equations,
\begin{equation}\label{abcd eqs}
\left\{
\begin{aligned}
0=&r\partial_r^2a-3\partial_ra-q^2r^3\partial_r c-3 i \omega r^2-q^2 r,\\
0=&r\partial_r^2b-3\partial_r b+ \frac{1}{3}r^3\partial_r c -\frac{2}{3} r^3 q^2\partial_r d -r,\\
0=&(r^7-r^3)\partial_r^2 c+(5r^6-r^2)\partial_r c - 2i \omega r^5\partial_r c\\
&-r\partial_r a + a- 3i \omega r^4 c+ 3r^4- i\omega r^3,\\
0=&(r^7-r^3)\partial_r^2 d+(5r^6-r^2)\partial_r d-2i \omega r^5\partial_r d\\
&+\frac{q^2}{3} r^3d-3i\omega r^4 d + 2b-2r\partial_r b-\frac{2}{3}r^3 c.
\end{aligned}
\right.
\end{equation}
The temperature is normalized to $\pi T=1$, so all momenta should be understood as  dimensionless: $\omega/(\pi T)$ and $q_i/(\pi T)$.

To find $T_{\mu\nu}$, we  consider  large $r$ behavior for the metric.
Near $r=\infty$, detailed analysis of Eqs.~(\ref{abcd eqs}) and~(\ref{k eq}) plus the boundary conditions~(\ref{AdS constraint}) and~(\ref{frame convention})
reveal
\begin{equation} \label{large r}
\left\{
\begin{aligned}
k(r)=&\frac{2}{3}\left(r^3+i\omega r^2\right)\partial\beta+ \mathcal{O}\left(\frac{1}{r^2}\right),\\
j_{i}(r)=&-i\omega r^3\beta_i-\frac{1}{3}r^2\partial_i\partial\beta + \mathcal{O}\left(\frac{1}{r}\right),\\
\alpha_{ij}(r)=&\left(\frac{2}{r}-\frac{\eta(\omega,q^2)}{4r^4}\right)\sigma_{ij}\\
&-\frac{\zeta(\omega,q^2)}{4r^4}\pi_{ij}+ \mathcal{O}\left(\frac{1}{r^5}\right),
\end{aligned} \right.
\end{equation}
where precise forms of $\eta$ and $\zeta$ will be determined via solving Eqs.~(\ref{abcd eqs}). The large $r$ behavior~(\ref{large r}) for the metric is related o the stress tensor of the boundary theory,
\begin{equation}
T_{\mu\nu}=T_{\mu\nu}^{\text{Ideal}}+T_{\mu\nu}^{\text{Diss}},
\end{equation}
where the ideal part $T_{\mu\nu}^{\text{Ideal}}$ is  $\frac{1}{{\bf b}^4}(\eta_{\mu\nu}+4u_{\mu}u_{\nu})$, which is linearized to
\begin{equation}
\begin{split}
T_{00}^{\text{Ideal}}&=3(1-4{\bf b}_1),~~~T_{0i}^{\text{Ideal}}=-4\beta_i,\\
T_{ij}^{\text{Ideal}}&=\delta_{ij}\left(1-4{\bf b}_1\right).
\end{split}
\end{equation}
The dissipative part $T_{\mu\nu}^{\text{Diss}}$ is nonzero only for  spatial components,
\begin{equation}
T_{ij}^{\text{Diss}}=-\left[\eta(\omega,q^2)\sigma_{ij} + \zeta(\omega,q^2)\pi_{ij}\right],
\end{equation}
where $\eta(\omega,q^2)$ is the generalized viscosity function proposed in Ref.~\cite{0905.4069} and $\zeta(\omega,q^2)$ is a new viscosity function emerging starting from the third order\footnote{In Ref.~\cite{0905.4069}, $\zeta$  was apparently  incorrectly argued to be zero.}. Eqs.~(\ref{abcd eqs}) are the main equations of this paper, which could be viewed as exact RG flow equations for the viscosity functions.

Generalized Navier-Stokes equations can be derived by focusing on the remaining Einstein equations.
More specifically, the large $r$ limits of $r^2f(r)E_{vr}+E_{vv}=0$ and $r^2f(r)E_{ri}+E_{vi}=0$ result in
\begin{equation}\label{ns}
\begin{split}
\partial_v {\bf b}_1=&\frac{1}{3}\partial\beta,\\
\partial_i {\bf b}_1=&\partial_{v}\beta_i -\frac{\eta(\partial_v,\partial^2)}{24}\left(\partial_i\partial\beta+ 3\partial^2\beta_i\right)\\
&-\frac{\zeta(\partial_v,\partial^2)}{6}\partial^2\partial_i\partial\beta.
\end{split}
\end{equation}
Fully consistently, equations~(\ref{ns}) can be shown to be equivalent to the conservation law $\partial^{\mu}T_{\mu\nu}=0$. Our task of deriving fluid dynamics from gravity is mathematically reduced to the boundary value problem of ordinary differential equations~(\ref{abcd eqs}).

We first perturbatively solve Eqs.~(\ref{abcd eqs}) by assuming $\omega$ and $q$ to be small. This procedure is equivalent to  the usual derivative expansion. We present the final results,
\begin{equation}\label{analytical}
\begin{split}
\eta(\omega,q^2)&=2+(2-\ln 2)i\omega-\frac{1}{4}q^2 - \frac{1}{24}\left[6\pi-\pi^2 \right.\\
&~~~\left.+12\left(2-3\ln 2+\ln^2 2\right)\right]\omega^2+ \cdots,\\
\zeta(\omega,q^2)&=\frac{1}{12}\left(5-\pi-2\ln 2\right)+\cdots,
\end{split}
\end{equation}
where, within our normalization, the first term in $\eta$ corresponds to $\eta/s=1/(4\pi)$; the second term in $\eta$ is the relaxation time \cite{hep-th/0703243,0712.2451,0712.2456}. The remaining two terms in~(\ref{analytical}) are  new third order transport coefficients.

\begin{figure}[h]
\includegraphics[scale=0.47]{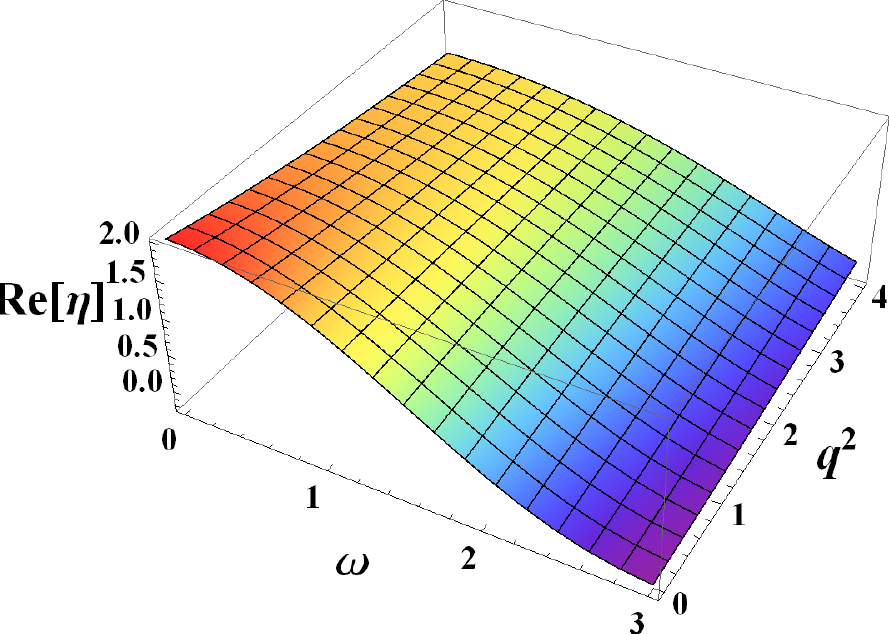}
\includegraphics[scale=0.47]{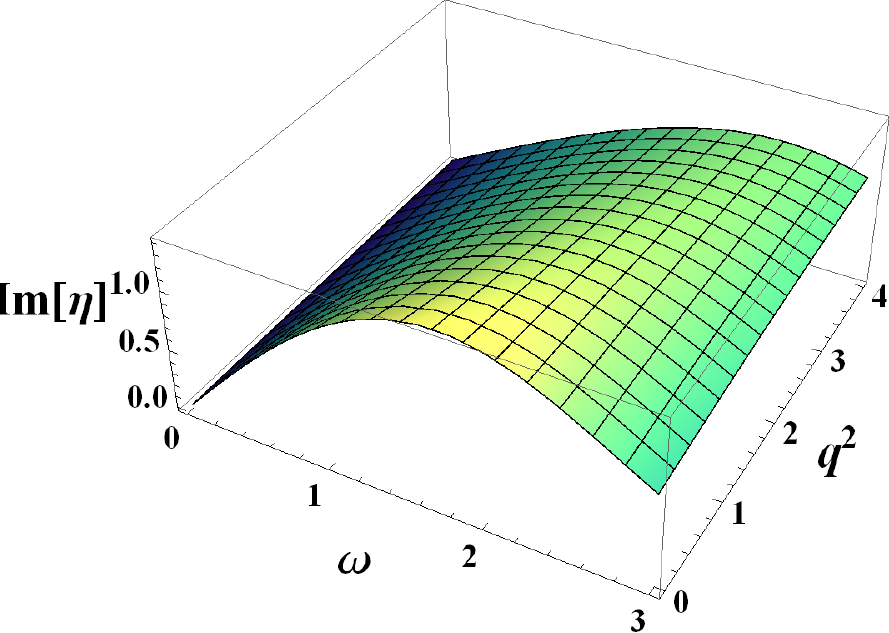}
\includegraphics[scale=0.47]{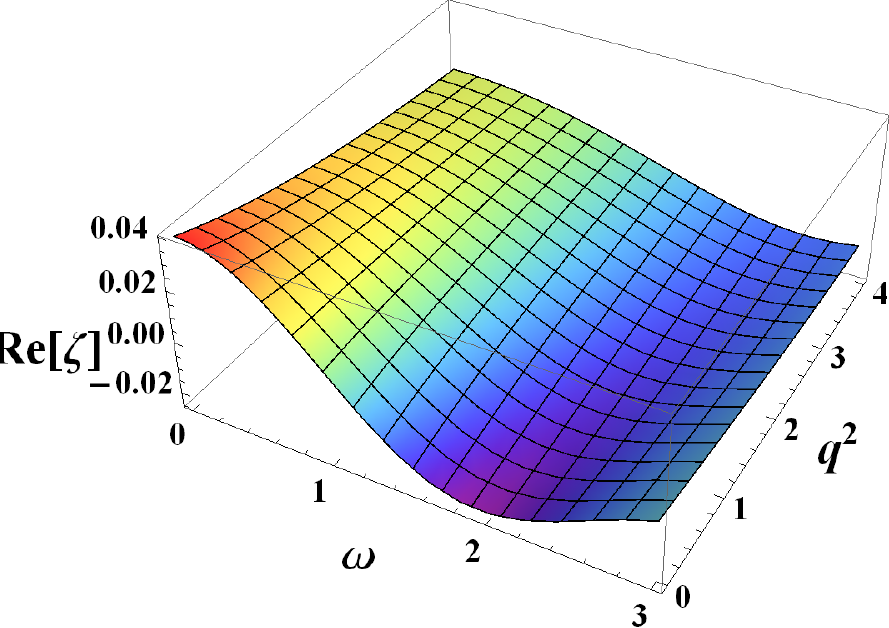}
\includegraphics[scale=0.47]{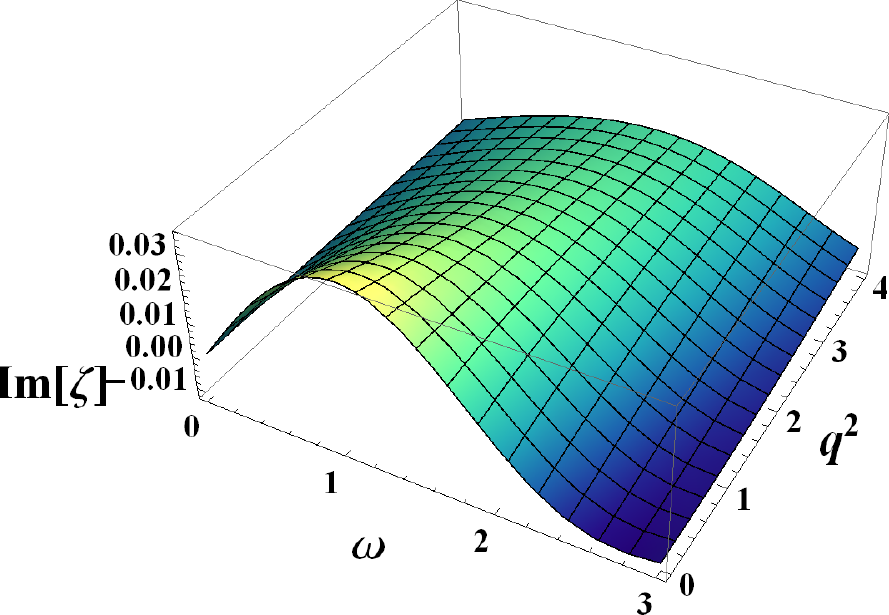}
\caption{Transport coefficient functions $\eta(\omega,q^2)$ and $\zeta(\omega,q^2)$ as functions of $\omega$ and $q^2$.}
\label{figure}
\end{figure}

To include all orders of boundary derivatives in $T_{\mu\nu}$, we now solve Eqs.~(\ref{abcd eqs}) numerically. Since we have to impose boundary conditions  both at the horizon and asymptotic infinity, we resort to a shooting technique. We first find  regular series solutions to Eqs.~(\ref{abcd eqs}) near $r=1$, with $6$ expansion coefficients unknown. The regular solutions near $r=1$ enable us to evolve Eqs.~(\ref{abcd eqs}) to $r=\infty$. The requirements~(\ref{AdS constraint}) and~(\ref{frame convention}) then completely fix these $6$  coefficients. We show our numerical results for the transport coefficient functions in FIG.~\ref{figure}.

Two features of FIG.~\ref{figure} are worth mentioning. In the hydrodynamic regime $\omega,q\ll 1$, the higher order terms have the effect of reducing the real parts of $\eta$ and $\zeta$, as already noticed in \cite{0704.1647}. A decrease in the effective viscosity affects dispersion relations in the fluid, such as sound waves. It was argued in \cite{0704.1647} that these finite momenta effects could be responsible for the extra low viscosity observed in plasmas produced in heavy ion collisions. We anticipate that both viscosity functions vanish at very large momenta as seen in FIG.~\ref{figure}.  This behavior is important for reliable discussion of early times in heavy ion collisions, thermalization and entropy production \cite{1203.0755,1108.3972}.
The second point is that, as far as the absolute values of $\eta$ and $\zeta$ are concerned, $\zeta$ is highly suppressed. Therefore, it looks reasonable to ignore $\zeta$ in  construction of an improved hydrodynamic model in the spirit of Ref.~\cite{0905.4069}.
\section{Conclusion}
We determined the linearized energy stress tensor of $\mathcal{N}=4$ super-Yang-Mills theory at strong coupling using the fluid/gravity correspondence. We obtained closed linear RG flow equations for the viscosity functions. Intriguingly, an analogous RG flow equation for conductivity derived in \cite{0809.3808}  is  nonlinear. We also derived the generalized Navier-Stokes equations for the dual fluid and checked the consistency of our formalism. To third order in derivative expansion, we analytically computed the stress tensor for the dual fluid. We summarized our results for the viscosity functions including all order derivative terms in FIG.~\ref{figure}. While our results on the stress tensor are exact even far beyond the hydrodynamic limit of small momenta, we obviously do not recover the entire UV physics, but only part of the dynamics related to the energy-momentum conservation.

We will report more details about this work in a forthcoming expanded publication \cite{forthcoming}.
\section{Acknowledgments}
YB would like to express his gratitude to Yun-Long Zhang for numerous discussions on fluid/gravity correspondence and  to  Jiajun Ma for useful discussion on  numerical calculations. ML thanks Edward Shuryak for early collaborative works that lead to this project.
This work was supported by the ISRAELI SCIENCE FOUNDATION grant \#87277111, BSF grant \#012124, and the Council for Higher Education of Israel under the PBC Program of Fellowships for Outstanding Post-doctoral Researchers from China and India (2013-2014).

\providecommand{\href}[2]{#2}\begingroup\raggedright\endgroup


\end{document}